\documentclass[12pt]{article}\usepackage[hyperfootnotes=false]{hyperref}
\usepackage{epsfig}
\usepackage{float}
\usepackage{empheq}
\usepackage{bbold}

\usepackage[utf8]{inputenc}
\usepackage{amsmath}

\usepackage{caption}

\usepackage{amsmath}
\usepackage{amssymb}
\usepackage{graphicx}
\newlength{\abstractwidth}
\renewcommand{\thefootnote}{\fnsymbol{footnote}}
\renewcommand{\thanks}[1]{\footnote{#1}} 
\newcommand{\starttext}{
\setcounter{footnote}{0}
\renewcommand{\thefootnote}{\arabic{footnote}}}

\newcommand{\be}{\begin{equation}}
\newcommand{\bea}{\begin{eqnarray}}
\newcommand{\eea}{\end{eqnarray}}
\newcommand{\beq}{\begin{equation}}
\newcommand{\ee}{\end{equation}}

\newcommand*\widefbox[1]{\fbox{\hspace{2em}#1\hspace{2em}}}

\def\dsp.{de Sitter space.}
\def\eq{&=&}

\def\la{\langle}
\def\ra{\rangle}
\def\simleq{\; \raise0.3ex\hbox{$<$\kern-0.75em
\raise-1.1ex\hbox{$\sim$}}\; }
\def\simgeq{\; \raise0.3ex\hbox{$>$\kern-0.75em
\raise-1.1ex\hbox{$\sim$}}\; }

\def\bi{\begin{itemize}}
\def\ei{\end{itemize}}

\def\CC{{\cal{C}}}

\def\CI{{\cal{I}}}

\def\CO{{\cal{O}}}

\def\CR{{\cal{R}}}

\def\bx{{\bar{\chi}}}

\def\Tr{\rm Tr \it}
\def\bsub{ \begin{subequations}
\begin{empheq}[box=\widefbox]{align}  }
\def\esub{ \end{empheq}
\end{subequations}}

\def\1{\(  \mathbb{1} \)}

  \def\bn{\bigskip \noindent}

    \def\dk{${\rm DSSYK_{\infty}}$}

\makeatletter
\g@addto@macro\normalsize{%
  \setlength\abovedisplayskip{10pt}
  \setlength\belowdisplayskip{20pt}
  \setlength\abovedisplayshortskip{10pt}
  \setlength\belowdisplayshortskip{20pt}
}
\makeatother

\usepackage{color}

\begin{document}


\begin{titlepage}


 \rightline{}
\bigskip
\bigskip\bigskip\bigskip\bigskip
\bigskip


\centerline{\Large \bf {A Parodox and its Resolution Illustrate  }} 

\bn

\centerline{\Large \bf { Principles of de Sitter Holography.}}

\bn



\bigskip
\begin{center}
	\bf   Leonard Susskind \rm

\bigskip

 Stanford Institute for Theoretical Physics and Department of Physics, \\
Stanford University,
Stanford, CA 94305-4060, USA \\ 

\bn

and Google, \\
Mountain View, CA

\end{center}

\bn


\begin{abstract}
Semiclassical gravity and the holographic description of the static patch of de Sitter space appear to disagree about properties of correlation functions. Certain 
holographic correlation  functions are necessarily real whereas their semiclassical counterparts have both real and imaginary parts.
The resolution of this apparent contradiction  involves the fact that time-reversal is a gauge symmetry in de Sitter space---a point made by Harlow and Ooguri--- and  the need for an observer (or quantum reference frame) as advocated by   Chandrasekaran, Longo, Penington, and Witten.

\end{abstract}

\end{titlepage}



 \rightline{}
\bigskip
\bigskip\bigskip\bigskip\bigskip
\bigskip

\starttext \baselineskip=17.63pt \setcounter{footnote}{0}

\tableofcontents

\Large

\section{Principles of de Sitter Holography}

My purpose in this note is to illustrate some basic principles of de Sitter holography by formulating and resolving a paradox which arises when these principles are combined. In order to be as clear as possible I will begin by first laying out the principles in a list.
\begin{enumerate}
\item The holographic principle may be applied to a static patch of de Sitter space. The holographic degrees of freedom  reside at the boundary of the static patch---that is to say at the (stretched) horizon. Reconstruction of the bulk (the static patch) should be done in terms of correlations of these stretched horizon degrees of freedom.
\item In the limit of very large horizon area the static patch may be described by the semiclassical equations of motion. In particular the approximation of quantum field theory in a fixed de Sitter background is accurate as  long as the energies involved are not large enough to create significant back reaction on the geometry. 
\item
Following \cite{Chandrasekaran:2022cip}
 the density matrix of the static patch is maximally mixed which means that it is proportional to the identity operator. The de Sitter entropy is equal to the  number of qubits comprising the system times $\log{2}.$ The expectation value of any operator supported in the static patch is given by
\be 
\la \CO \ra = \Tr  \ \CO
\label{tr=exp}
\ee
where $\Tr$ indicates the normalized trace, normalized so that the trace of the identity operator is unity.

An equivalent assumption is that the  formal input  temperature in the Boltzmann distribution is infinite. The finite Hawking temperature, or tomperature, is a derived or emergent quantity \cite{Lin:2022nss}.
\end{enumerate}
Later, again following \cite{Chandrasekaran:2022cip}  we will add another principle concerning the existence of an observer at the pode, i.e., the center of the static patch.

\section{The Paradox}

\begin{figure}[H]
\begin{center}
\includegraphics[scale=.4]{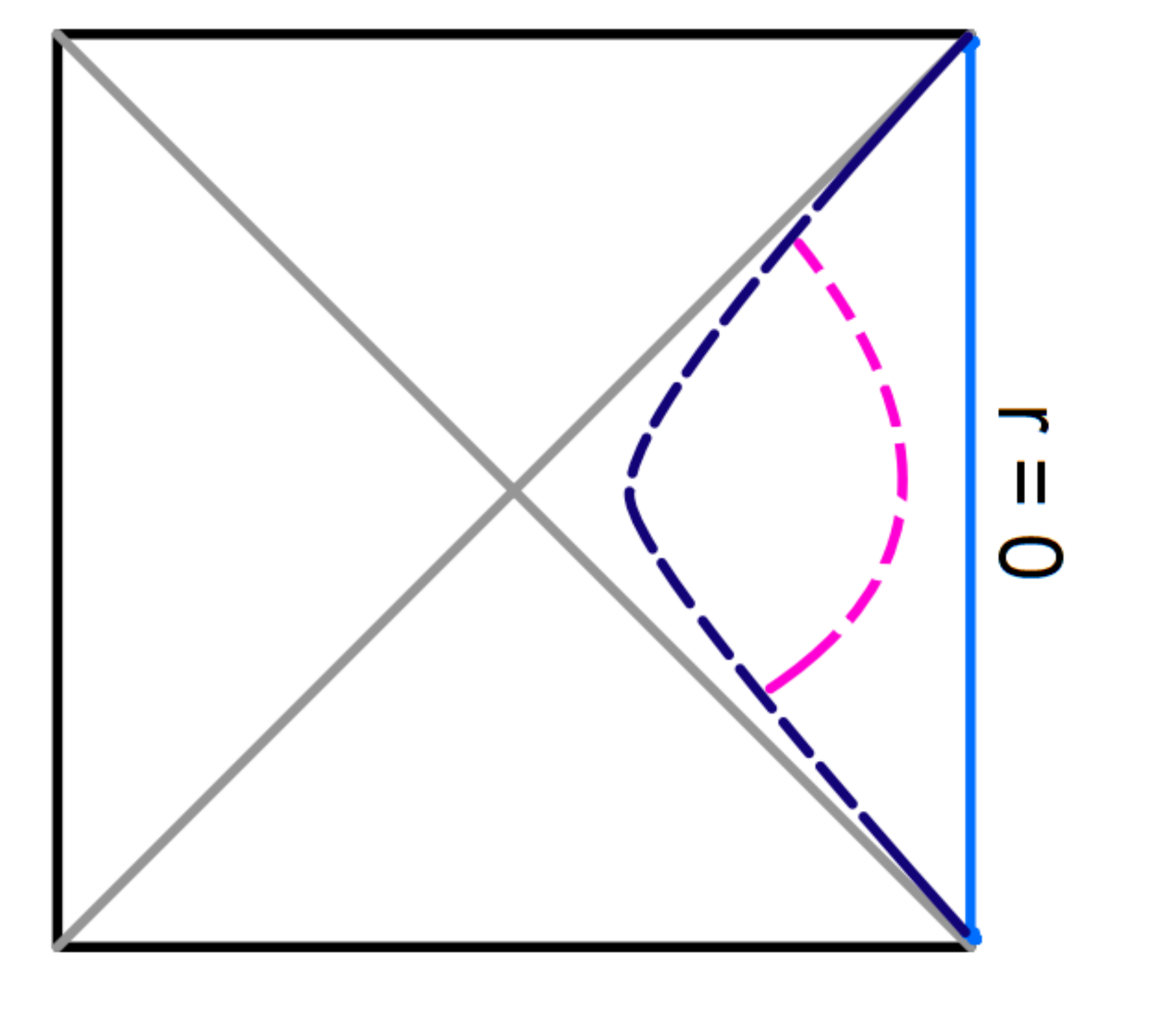}
\caption{Penrose diagram for de Sitter space. The dashed black curve is the stretched horizon of the right-side static patch. The dashed red curve shows a particle being emitted and absorbed by the stretched horizon.}
\label{correlation}
\end{center}
\end{figure}

Let's consider a two-point correlation function of bulk fields; the two points being  located on the stretched horizon as in figure \ref{correlation}.
\bn
 Intuitively the correlation function is the amplitude for a particle emitted from the stretched horizon at $t_1$ to be reabsorbed at $t_2.$ On the holographic side the amplitude is represented by the expectation value of the product of holographic operators. A concrete example of such holographic operators drawn from the complex (or charged) \dk-de Sitter duality would be the  $SU(N)$-singlet  massive photon operator \cite{Susskind:2023hnj} which is present when the $U(1)$ symmetry is slightly broken. I'll denote it by $A$,
\be 
A = \sum_i \bx_i  \chi_i.
\label{A=xx}
\ee
The correlation function is,
\be 
 \la A(t_1) A(t_2) \ra.
\label{C=AA}
\ee

In the semiclassical theory it can be calculated in the fixed background approximation. The  important point for this paper is that it has both a real and imaginary part,
\bea
 \la A(t_1) A(t_2) \ra\eq  \CR +\CI \cr \cr
\CR \eq \text{Re} \la A(t_1) A(t_2) \ra \cr \cr
\CI \eq \text{Im} \la A(t_1) A(t_2) \ra
\label{R and I}
\eea 
The real and imaginary parts are given in terms of anticommutators and commutators,
\bea 
\CR \eq \frac{1}{2} \la \ \{ A(t_1) , A(t_2)    \} \  \ra  \cr \cr
\CI \eq  \frac{1}{2} \la \ [   A(t_1)  , A(t_2)   ] \  \ra .
\label{comandacom}
\eea
In the semiclassical fixed-background approximation both are non-zero.

But now apply  \eqref{tr=exp} to the imaginary part,
\be 
\CI = \frac{1}{2} \Tr [   A(t_1)  , A(t_2)   ] =0.
\label{C=0}
\ee
The vanishing of $\CI$ follows from the fact that the trace of a commutator is zero. Thus we see a contradiction between the semiclassical fixed-background approximation and the holographic description in which the density matrix is maximally mixed.

This problem is not special to  \dk. It occurs in any holographic description of a static patch of de Sitter space that satisfies the three principles stated above.

\section{T-reversal as a Gauge Symmetry}
The same conclusion can be seen from an observation due to Harlow and Ooguri \cite{Harlow:2018tng}. These authors argue that in general, global symmetries including discrete symmetries, of the holographic boundary theory (in the case of de Sitter space the horizon theory) are gauge symmetries of the bulk theory. In particular they suggest that this applies to time-reversal\footnote{In this context time reversal means any symmetry that includes a reversal of the direction of time. In AdS/CFT such a symmetry is guaranteed by the CPT theorem applied to the boundary CFT. In charged SYK time-reversal may include charge conjugation.}.

The implications for de Sitter space of time-reversal-as-a-gauge-symmetry are: the state of the static patch must be invariant; and all operators in the algebra of observables must be time-reversal invariant. 

Now consider the commutator $\CC(t) =[A(-t), A(t)].$ It satisfies,
\bea
\CC(t) \eq [A(-t),A(t)] \cr \cr
\eq -[[A(t),A(-t)]     \cr \cr
\eq -\CC (-t).
\label{odd}
\eea
We see that the commutator is odd under t-reversal and cannot belong to the algebra of gauge invariant observables, with the trivial exception that it may be identically zero.
This argument seems very general.

\section{Resolution}
The resolution of the paradox involves another observation in \cite{Chandrasekaran:2022cip}, namely that to recover the usual semiclassical limit it is necessary to assume the existence of an observer in de Sitter space.  An observer means two things: the subsystem comprising the observer should be rich enough to keep track of the results of its interactions with other degrees of freedom in the static patch; and, most important for our purposes, the observer subsystem should include a clock. The clock provides a reference to define the zero of time. Without the clock correlation functions like $\CC(t_1,t_2)$ could not be defined. At best an integrated version,
  \be
\lim_{T\to \infty}  \frac{1}{2T} \int_{-T}^T  \CC(t_1+t,t_2+t )dt
\label{intversion}
\ee
would be gauge invariant. 

The existence of  the  clock is best understood as a gauge fixing procedure which I will describe in a separate publication. The main point of \cite{Chandrasekaran:2022cip} is that such a gauge fixing is necessary to recover the standard semiclassical theory in fixed background.

An analogous  procedure is necessary to fix the time-reversal gauge symmetry. If the boundary theory is time-reversal symmetric then the existence of a clock which ``runs forward" in time implies the possible existence of a  clock that runs backward. As an example a heavy coordinate $X$ with Hamiltonian,
\bea
h(P) &=& \frac{P^2}{2m} \cr \cr
[X,P] \eq i \hbar
\label{clock h}
\eea
may be used as a clock variable if the initial wave function is composed of positive momenta in a narrow enough range that the  function $h(P)$ is effectively linear. Then $X$ will  increase linearly with the formal time of the boundary theory. 

But the same system allows for a clock which runs backward. By considering a wave packet composed of negative values of $P$ over a similar range, the clock coordinate will run backward toward increasing values of $-X.$ Both types of clocks are equally likely to occur as fluctuations in de Sitter space.

Suppose for a moment that the observer's clock is time-reversal invariant. This could occur because the wave function is symmetric under $P\to -P,$ i.e., a superposition of right and left moving wave packets.  In that case the gauge fixing would be partial and would result in an average over forward and backward running clocks. The result is    that the expectation value of time-reversal odd operators would vanish, including the expectation value of $\CC$ and therefore the imaginary part of the correlation function $\la A, A \ra.$  On the other hand if the observer's clock is forward-going (or backward-going)  then the t-reversal gauge symmetry would be fixed and t-odd operators would not generally vanish.

In the appendix to this note I further elaborate on the relation between the existence of observers and gauge fixing.

\section{Quantum Reference Frames (QRFs)}

As noted in \cite{Chandrasekaran:2022cip} one cannot just say that the observer has a clock. A clock must consist of a superposition of energy levels but the constraint of gauge invariance requires the total energy to be zero in the gauge-invariant description. The quantum state of the clock cannot be pure but must
 be entangled with other degrees of freedom so as to fix the total energy to zero. Reference \cite{Chandrasekaran:2022cip} describes this situation in terms of  von Neumann algebras.  An alternative
 framework for this purpose is the language of quantum reference frames (QRFs) initiated in the 1960's \cite{Aharonov:1967zza}. (For a more recent study of quantum reference frames see \cite{Bartlett:2006tzx}.)
 
 According to the theory of QRFs a reference frame is a physical object that has properties such as energy, momentum, charge, or whatever else is relevant. Another subsystem, such as a particle,  is not to be thought of as localized in some abstract sense; it may be localized only relative to the QRF. This means that it is entangled with the QRF in a specific manner in which the subsystems ``share" a conserved quantity. The QRF viewpoint is particularly relevant for a closed system like de Sitter space where gauge conditions require the conserved quantity to be zero.
 
  Let's consider the case of spatial localization. A particle can only be localized relative to a QRF if the two ``share" momentum; the total momentum being zero.
  Relative localization requires the overall state of the particle and QRF to have the form,
  \be  
  |\Psi \ra = \sum_p  e^{ipa}  \ \ |p\ra_{particle} \otimes  \  |-p\ra_{QRF}.
  \label{p-syn}
  \ee
  
  Then one says that the particle is localized at position $a$ relative to the QRF.
  
  Similarly for the case of a conserved charge $q$. A subsystem $s$ is localized in the conjugate phase $\theta$  if the state has the form,
    \be  
  |\Psi \ra = \sum_q  e^{iq\theta} \ \  |q\ra_{s} \otimes  \  |-q\ra_{QRF}.
  \label{qsyn}
  \ee
  
  If initially a particle is in a pure state it will be completely delocalized. If it subsequently interacts with the QRF the two may become entangled,  the particle becoming localized relative to the QRF.
  
  One may consider a pair of QRFs called $A$ and $B$. If the two frames are unentangled then they are delocalized with respect to one another. If a particle interacts with $A$ and becomes localized in the frame $A$ it will still be delocalized relative to $B.$  But if $A$ and $B$ are relatively localized with respect to one another:
  \be  
 | \Psi_{AB}\ra = \sum_p e^{ipa} \ \  |p\ra_A  \ \otimes \ |-p\ra_B 
  \label{ABsyn}
  \ee
  then a particle localized in $A$ will also be localized in $B.$
  
  Localization in time is similar. Suppose we have two clocks which are unentangled. The readings of the clocks will be uncorrelated. To correlate (synchronize) them  we must allow them to share energy,
  \be 
  |\Psi \ra = \sum_E |E\ra_A  \ \otimes \ |-E\ra_B
  \label{TFDinf}
  \ee
In the Fourier transformed time-basis,
\be 
|\Psi\ra = \int dt |t\ra_A \otimes \ |t\ra_B
\label{tsync}
\ee

Equation \eqref{TFDinf} defines the infinite temperature thermofield-double state. Eq. \eqref{tsync} tells us that the two members of an infinite temperature thermofield-double are perfectly synchronized  QRFs for time (clocks).

In anti de Sitter space QRFs do not play much of a role. The frames of reference---clocks, measuring rods, etc---may be regarded as being outside the system in the ``lab" beyond the boundary. But in de Sitter space everything, including the observers and the reference frames they carry with them, are part of the system. Moreover since the symmetries are gauged, the total value of the conserved charges is zero. In this situation the QRFs are an indispensable part of the description.

Similar considerations apply to discrete symmetries including time-reversal or CPT. The observer's clocks may run forward or backward relative to the (unphysical) mathematical time conjugate to the Hamiltonian. 

Let us suppose that there are a pair of clocks in de Sitter space, one in each of a pair of static patches. Each clock can run backward or forward. If the state is time-reversal-invariant (as required by gauge invariance) then it has the form,
\be 
|\Psi \ra = |\text{forward}\ra_A  \otimes  |\text{forward}\ra_B + |\text{backward} \ra_{A} \otimes  |\text{backward}\ra_B
\label{ff+bb}
\ee  
In this gauge-invariant description any time-reversal odd operator will vanish because it will be averaged over forward and backward going time.

But in the gauge-fixed version where we pick one of the two branches---say the forward branch---time-reversal odd operators do not generally vanish.

 This resolves the paradox, but more importantly it illustrates the interplay between the various principles of de Sitter holography and the need for observers, or quantum reference frames. In a separate publication I will describe the construction of an observer in the concrete example of \dk.

\section*{Acknowledgements}
I thank Daniel Harlow for explaining  that time-reversal (or CPT more generally) is a gauge symmetry in de Sitter space.  That discussion was one of the motivations for   this note. I am also grateful to Edward Witten for helpful remarks.

\section*{Appendix}
The theory of a static patch is all about fluctuations \cite{Susskind:2021omt}. Even the existence of an observer is a fluctuation. To see this suppose that an observer is known to exist at the pode  at $t=0.$ If we run the theory into the future the observer will eventually evaporate and disappear into the stretched horizon. But it is also true that if we run the state into the past the observer will also disappear into the stretched horizon. The entire history of the observer consists of a fluctuation in which the observer is emitted into the static patch and eventually is reabsorbed  into the horizon degrees of freedom.

It is well known that the presence of an observer or any other fluctuation will decrease the total entropy of the static patch, say by amount $\Delta.$ The probability for the fluctuation is $P=e^{-\Delta}$ which is typically very small but non-zero. In a time-reversal invariant theory forward and backward going clocks will occur with equal probability.

Given enough time a clock or any other fluctuation will occur with probability one. That is true for both forward-going and backward-going clocks. This is the reason for the vanishing of time-reversal odd expectation values.

On the other hand symmetries like time-translation and time-reversal are gauge symmetries and as with all gauge symmetries the gauge can be fixed. As an example, in QED the expectation value of the space-component of the vector  potential will be zero in a gauge-invariant formulation, even in the presence of a current carrying wire. However gauge-fixing to the Coulomb gauge breaks the gauge invariance and leads to the usual non-zero expression for the vector potential of a current. This is the most familiar example of how an quantity can be zero for all gauge-invariant states but non-zero in the gauge-fixed semiclassical version of the theory. The main difference between the electromagnetic  and  de Sitter examples is that in a quantum theory of gravity the gauge-fixing has to be relative to a physical quantum reference frame. 

The gauge-fixing for time-translation invariance is accomplished by first realizing that there is unit  probability  that at some point there will be a fluctuation leading to the  formation of a clock; and then fixing the zero of time so that it corresponds to the clock's formation. Once that is done, the equations will no longer appear time-translation invariant \cite{Chandrasekaran:2022cip}.

Exactly the same thing can be said for time-reversal. One further fixes the gauge
by requiring not only the existence of a clock at $t=0$ but that the clock is forward (or backward) going. The equations of the gauge-fixed theory will no longer be time-reversal invariant,  and in the gauge-fixed theory the imaginary part of correlation functions will not vanish.

\end{document}